# Interpreting neural computations by examining intrinsic and embedding dimensionality of neural activity

Mehrdad Jazayeri[1], Srdjan Ostojic[2]

## Summary

The ongoing exponential rise in recording capacity calls for new approaches for analysing and interpreting neural data. Effective dimensionality has emerged as an important property of neural activity across populations of neurons, yet different studies rely on different definitions and interpretations of this quantity. Here we focus on intrinsic and embedding dimensionality, and discuss how they might reveal computational principles from data. Reviewing recent works, we propose that the intrinsic dimensionality reflects information about the latent variables encoded in collective activity, while embedding dimensionality reveals the manner in which this information is processed. We conclude by highlighting the role of network models as an ideal substrate for testing more specifically various hypotheses on the computational principles reflected through intrinsic and embedding dimensionality.

## Highlights

-    Dimensionality has emerged as a key concept for describing collective neural activity
-    Characterizing intrinsic and embedding dimensionality provides a guiding principle for interpreting underlying computations
-    Intrinsic dimensionality reflects the nature of the information encoded in collective activity
-    Embedding dimensionality describes how information is processed by neural circuits
-    Network models provide a testing ground for the computational roles of intrinsic and embedding dimensionality

## Addresses

[1]McGovern Institute for Brain Research, Department of Brain & Cognitive Sciences,
Massachusetts Institute of Technology, Cambridge, Massachusetts 02139, USA

[2]Laboratoire de Neurosciences Cognitives, INSERM U960, École Normale Supérieure - PSL Research University, 75005 Paris, France

**Introduction**

Ongoing breakthroughs in tools and technologies for neural recording have sustained an exponential rise in the number of neurons that can be simultaneously recorded in behaving animals (Urai et al. 2021; Stevenson and Kording 2011). The number of neurons we can record from in a brain area however, usually far exceeds the number of variables relevant to any given behavioral task. For example, in the primary motor cortex (M1), millions of neurons interact to control just a few muscles involved in a movement [2]. Similarly, interactions between large populations of neurons in the frontal and parietal cortex are thought to support simple decision-making tasks that involve just a few relevant variables [3–5]. This mismatch makes signals across neurons highly redundant and raises the question of how the population activity represents task-relevant variables [6–8].

To quantify this redundancy, and identify shared components of collective dynamics that reflect task variables, the effective dimensionality of population activity has emerged as a key concept. A central difficulty however is that different studies rely on different notions of dimensionality, so that a unifying picture, and even a common terminology, are currently lacking. Here we focus on the distinction between two types of dimensionality of neural activity, which we refer to as *intrinsic* and *embedding* dimensions (see Box 1), and propose that they provide guiding principles for interpreting neural computations. Reviewing recent experimental studies, we argue that intrinsic dimensionality in different brain areas reflects the nature of the information encoded across neurons in those areas, while the embedding dimensionality is a signature of the manner in which that information is processed. Systematically contrasting intrinsic and embedding dimensionality in different brain areas and behavioral settings is therefore a potentially powerful approach for generating hypotheses on neural computations. We finally discuss how the two types of dimensionality can be controlled within network models to implement specific computational hypotheses and produce further experimentally testable predictions.

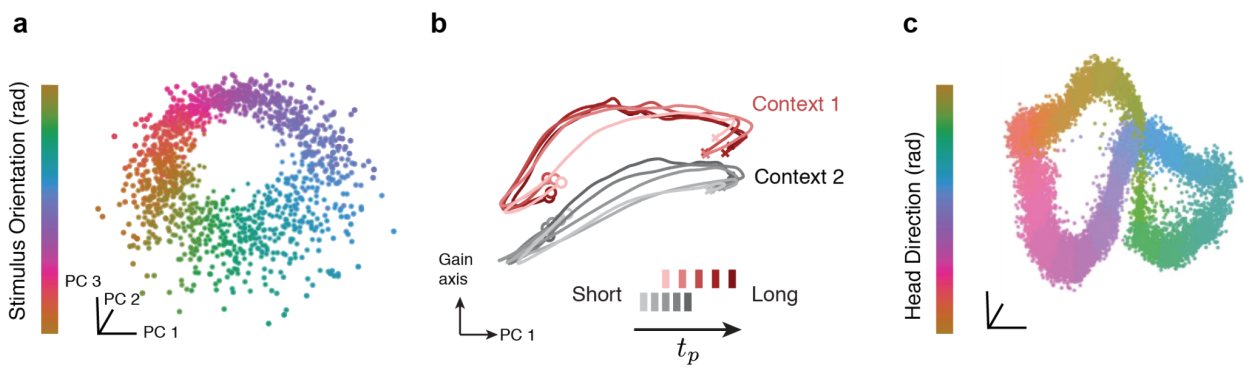

**Figure 1.** Non-linear organization of neural activity. **a** Population activity in the mouse primary visual cortex in response to gratings of different orientations indicated in color. Projection on the first three principal components of data from [9]. **b** Population activity in the macaque dorsomedial frontal cortex reflecting flexible motor planning in a context-dependent timing task [10]. The animal had to reproduce different time intervals in two contexts. In one context, the produced interval ($t_p$) had to match the sample, and in the other, $t_p$ had to be 50% longer than the sample (gain=1 or 1.5, respectively). Different traces show DMFC population activity associated with the two contexts (gray and red) for different samples. Despite the fact that $t_p$ distributions for the two contexts were part of a continuum, neural responses were separated in a highly nonlinear fashion along a "Gain axis" distinguishing between the two contexts. **c** Population activity in the rat head-direction system [11]. Colors represent a computationally-inferred angular latent variable that maps directly onto head orientation. Panel **a** courtesy of NeuroMatch Academy [12] Pod 001 (Daniela Buchwald, Agustina Frechou, Habiba Noamany, Antonio Ortega).

**Dimensionality of Neural Activity**

The instantaneous firing rates across a sample of *N* neurons can be conceptualized as a point, or a *neural state*, in an *N*-dimensional Euclidean coordinate system known as the *state-space* where each axis represents the activity of one neuron. Within this framework, one can define the dimensionality in different ways. First is the dimensionality of the entire state space (*N*), which we refer to as *ambient dimensionality*. Ambient dimensionsionality corresponds to the space of all possible neural states and thus does not reflect anything about the structure of neural activity.

Task-relevant computations usually depend on structured neural activity whose dimensionality is lower than that of the state space [13]. Quantifying this dimensionality is important as it bears information about the underlying variables and computations [11,13–16]. One common strategy to quantify this reduced dimensionality is to apply principal component analysis (PCA) to the neural data, and determine the number of patterns of activity that account for a given fraction of variance. The specific criterion for the amount of explained variance differs across studies, a common threshold being 80% of the total variance. Here we refer to the number of Euclidean dimensions that are required to capture a given fraction of the structured activity as the *embedding dimensionality* of the data.

While the embedding dimensionality carries information about structure, because of an assumption of linearity, it does not in general correspond to the number of independent variables needed to describe the data (Fig. 1). For instance, a set of points organized along a ring that can be described by a single circular variable can be embedded in a state space with any number of dimensions (Box 1). More generally, we call the number of independent variables the *intrinsic dimension*, and we refer to each independent variable describing the neural activity as a *latent variable*.

Recent studies have combined simple linear methods such as PCA with more sophisticated nonlinear dimensionality reduction techniques to quantify both the intrinsic and embedding dimensionality of neural data providing insight into the nature of the underlying representations and computations [10,11,14,17–25]. For example, Chaudhuri et al. [11] analyzed neural activity associated with a population of head direction cells in the anterodorsal thalamic nucleus of mice during active foraging as well as rapid eye movement (REM) sleep. Using a novel nonlinear decoding strategy, they found that the population activity was organized around a ring structure with intrinsic dimensionality of 1, and that the position on the ring parametrically encoded animals' head direction. The ring, however, had a nontrivial geometry, much like a twisted elastic band, so that the embedding dimensionality was much larger than one.

In another study, Low et al. [14] analyzed neural activity in the hippocampus and entorhinal cortex of mice during navigation. Using a novel nonlinear manifold inference technique, they were able to show that a large proportion of variance in the population activity was constrained to a manifold with intrinsic dimensionality of approximately 3, and that two of the dimensions corresponded to the animals' navigational space. However, due to its nonlinear embedding in the state space, the manifold was associated with a much larger embedding dimensionality. Intuitively, it was as if the 2D plane of navigation was twisted and stretched before being embedded in the neural state space.

Estimating the embedding and intrinsic dimensions from noisy data is in general a challenging problem, and most methods appear to overestimate these quantities (Altan et al. 2020). Sidestepping this issue, in this review we focus on the key question of how latent variables and their embedding might relate to task variables and the underlying computations.

---

**Box 1: Representing neural data in terms of manifolds**

A premise of the analyses of dimensionality is that neural activity is restricted to one or several *manifolds* within the full state space.

A manifold is an idealized mathematical object that describes a continuous set of points. A manifold is characterized by two types of dimensions, which we refer to as the intrinsic, or non-linear dimension, and the embedding, or linear dimension. The ***intrinsic dimension*** is the minimal number of continuous variables needed to parametrize the manifold. The ***embedding dimension*** is in contrast a measure of the number of dimensions explored by the manifold within the ambient Euclidean space.

A simple example of a non-linear manifold is a set of points forming a ring. The intrinsic dimension of a ring is one, as all the points can be parametrized by a single continuous variable, an angle along the ring. The embedding dimension of a ring however depends on how much it is twisted within the ambient space. A flat ring is embedded in two dimensions, while a warped ring may be embedded in three or more dimensions, although it is still parametrized by a single angle. A spherical shell is a more complex example, where two parameters are needed to index every point, while the whole manifold may be embedded in three or more dimensions.

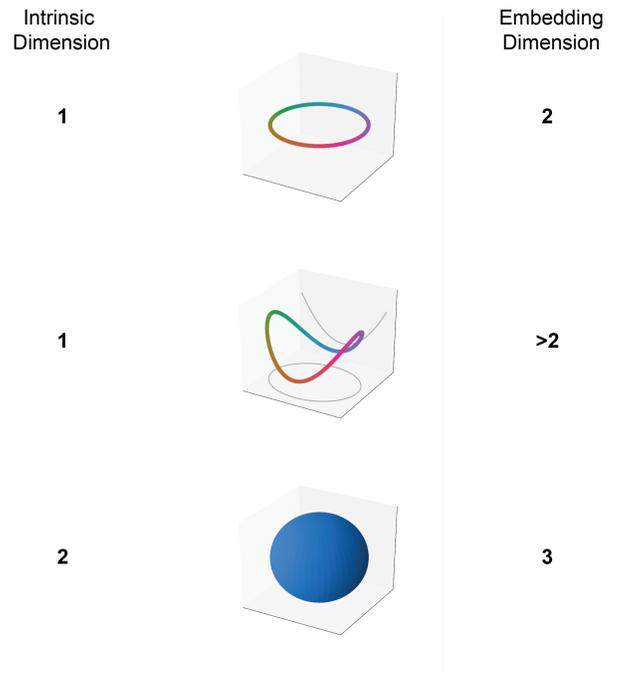

For an idealized manifold such as a ring, the intrinsic and embedding dimensions are unambiguously defined. In contrast, for noisy datasets such as neural recordings, estimating the dimensionality is a challenging problem. While a number of algorithms have been developed for non-linear dimensionality reduction, accurately estimating intrinsic and embedding dimensions from noisy data remains an active area of research [22,26].

---

**Computational factors influencing intrinsic dimensionality**

From a computational perspective, the intrinsic dimensionality of neural activity is largely determined by three sources of information: (1) incoming stimuli, (2) ongoing movements, and (3) the multitude of latent variables that characterize prior experiences and future expectations. This information, however, is not uniformly present in all brain areas. In early sensory areas, the intrinsic dimensionality is expected to be strongly associated with incoming stimuli. For example, responses to gratings in the primary visual cortex are organized along a circular variable corresponding to the grating orientation [9,18,21] (Fig 1a). Similarly, responses in the olfactory cortex reveal the latent organization of the chemical odour space [27]. The intrinsic dimensionality for more natural stimuli has not yet been systematically quantified, but it is likely that it would increase with the complexity of spatial [28] and/or temporal [29,30] dependencies.

On the other end of the spectrum are the neurons in the motor cortex and areas downstream that directly control muscle movements. Electrophysiology experiments and neural network modeling suggest that the dynamics of population activity in the motor cortex provide a potential basis set for outgoing muscle-like commands [7,31–35]. Remarkably, signals in the primary motor cortex seem to carry information about the upcoming movement without regard to higher order latent variables such as context [36] or even the movement after the next [37]. As such, the intrinsic dimensionality in the late stages of the motor system seem to be strongly tied to ongoing movements (Fig 1b).

On the path from early sensory to late motor areas, neural signals go through a web of recurrent cortical and subcortical circuits that carry abstract information about sensory variables such as magnitude [38], navigation variables such as head direction (Fig 1c) [11] or position [20,23,25] , and other latent variables including behavioral context [4,10,15,39], contents of working memory [40], beliefs about rules [41], attentional state [42,43] and other internal states such as arousal and motivation [44]. Estimating the intrinsic dimensionality associated with these latent variables is a daunting task. First, long term dependencies in the structure of data can cause dimensionality to be exceedingly high [45]. Second, for cognitive tasks of any appreciable complexity, the latent variables the brain uses to control behavior are not known. One loosely articulated idea, motivated by humans' remarkable generalization capacity, is that these latent variables are organized such that knowledge about past experiences can be applied flexibly to new situations. One leading hypothesis from cognitive neuroscience and machine learning is that the brain decomposes experiences into explanatory factors such as objects, events, causal relationships, and other regularities that could support inference, and planning in new situations [46,47]. However, how these explanatory variables might map onto intrinsic variables and determine their dimensionality are among the most important unresolved questions at the intersection of machine learning and neuroscience [48]. Nonetheless, recent work has begun to make progress in this direction by developing sophisticated nonlinear dimensionality reduction techniques and applying them to large-scale neural data [10,11,14,19,49], and by using artificial neural network models to generate testable hypotheses for how neural systems might emulate complex cognitive behaviors [16,50,51].

**Computational factors influencing embedding dimensionality**

While the intrinsic dimension of the non-linear structure of population activity reflects stimuli, movements, and other core latent variables, its embedding into the Euclidean state space determines how these variables are processed and communicated. Different hypotheses about the nature of information processing in the brain can be casted as different forms of embedding. For example, in a system in which individual neurons and neural circuits play distinct functional roles, the state space can be parsimoniously partitioned into subspaces spanned by subsets of neurons. In contrast, when distinct sources of information are processed by partially overlapping groups of neurons (i.e., "mixed selectivity" [52]), the corresponding embeddings are no longer aligned with the principal axes of the state space.

Although the general principles that govern embedding dimensionality are not known, several computationally inspired hypotheses have been proposed. One dominant hypothesis is that the information in any given brain area is extracted by other areas and peripheral systems through a linear readout. A linear readout scheme is a weighted average of the activity across neurons and has an intuitive geometric interpretation: it is a projection of the neural activity in the full Euclidean state space along a specific direction. One common strategy to evaluate this hypothesis is to quantify the degree to which a linear decoder can extract desired information from population activity in a brain area.

An extension of this hypothesis is that embedding across a population of neurons is organized such that different linear decoders can extract information about different task-relevant variables without interference. For example, movement and movement preparation signals in monkeys' motor cortex reside in orthogonal subspaces [53]. This organization could ensure that preparation does not trigger movement [54,55], and may help protect previous motor memories [56]. Similar principles might govern interactions across cortical areas, by confining to orthogonal subspaces information that is private to an area and information that is communicated [57,58]. Orthogonal subspaces have also been implicated in the auditory cortex, both for reducing interference between new stimuli and old stimuli [40], and to enhance the representation of behaviorally relevant stimuli during task engagement [59]. Finally, orthogonal subspaces are thought to enable neural representations to disentangle working memory from motor preparation [60], decision variables from confidence and contextual cues [4,10,24] or represent different behavioral states (Lanore et al. 2021). We note however, that despite strong enthusiasm, direct evidence that the brain processes and/or communicates information through orthogonal linear decoders is wanting.

In general, high-dimensional embeddings with more degrees of freedom can facilitate extraction of task-relevant variables without interference (Rigotti et al. 2013; Cayco-Gajic et al. 2017; Cayco-Gajic and Silver 2019; Litwin-Kumar et al. 2017; Lanore et al. 2021). However, embedding information in arbitrarily high dimensional subspaces can have adverse effects for generalization [64]. To improve generalization, embeddings have to be appropriately constrained to capture the structural relationships and inherent invariances in the environment [10,38,65]. A theory for the ensuing trade-offs has been recently developed for the case where the activity is organized in multiple disjoint manifolds corresponding to different categories [66] and applied to interpret representations in the visual system [67] and in deep networks [68]. It is also possible to organize information embedding such that certain linear projection reflect information in more abstract form and therefore enable generalization, while others enable finer discrimination [15].

The idea of linear decodability, orthogonal subspaces, and their desiderata have straightforward interpretations for simple experiments in early sensory and late motor systems. For example, assuming that a core function of the early visual system is to categorize objects, modern models of vision are built to make object categories linearly decodable [69]. Similarly, in the motor system, open-loop movements such as ballistic reaches are ultimately constrained by how the activity is projected onto motor neurons, and models based on this idea can emulate neural responses in the motor cortex [33]. The utility of linear decodability however, becomes less clear for intermediate stages of information processing in higher brain areas that carry information about latent variables that support flexible mental computations. While an experimenter may apply linear decoders to find information about a hypothesized latent variable in a certain brain area, there is no a priori reason to assume that the brain relies on such decoders. In these situations, the computational constraint that likely determines the embedding properties of latent variables is their controllability [70]; i.e., the ability to flexibly use the underlying representation in the context of different policies, plans and action [71,72]. While there has been growing interest in the general notion of controllability in neural networks [73], our current conception of this problem is at its infancy [74] and more work is needed to understand how control may impact embedding dimensionality.

**Network models as tools for testing hypotheses about intrinsic and embedding dimensionality**

Large-scale recordings of neural activity together with innovative analysis techniques have paved the way toward more accurate measurements of intrinsic and embedding dimensionality in different brain areas and in a wide range of tasks. However, without concrete computational hypotheses, it could be extremely challenging to interpret measures of dimensionality. A powerful substrate for instantiating specific hypotheses and generating testable predictions are neural network models that can produce a desired behavior in a given task [51,75] or emulate patterns of neural activity associated with a certain brain area [76–78]. Comparison of neural network models to brain activity can narrow the space of hypotheses [50], and detailed analysis of candidate models can give insight into the underlying computational mechanisms [79]. For example, recent modeling efforts have found that the structure of neural activity in a wide range of brain areas is tightly linked to task requirements [3,10,39,80–84]. Systematic characterization of the underlying intrinsic and embedding dimensionality in network models however remains an open issue, and here we review promising paths forward.

Similarly to cortical activity, network models transform incoming sensory latent variables into behaviorally relevant outputs through a set of intermediate, abstract latent variables. A central question is how network dynamics expand the intrinsic dimensionality of activity to generate additional, internal latent variables that represent abstracted information not explicitly present in immediate sensory inputs. Recurrent dynamics are a key candidate mechanism for this process, as they can integrate past information to extract long-term dependencies relevant for the task. Recent works have examined mechanistically how the interplay between feed-forward and recurrent connectivity determines the dimensionality of activity [16,85], and shapes the dynamics of the abstract latent variables that emerge from recurrent dynamics [86]. For instance, ring-like manifolds, on which activity is represented by a single angular latent variable, can emerge from only weak structure in the connectivity [16,86], while more complex non-linear manifolds can be generated with additional structure [87]. Such modelling frameworks allow for a direct control of intrinsic and embedding dimensionality, and provide a rich testing ground for examining hypotheses on their computational roles.

A particularly important question pertains to the number and nature of intrinsic latent variables needed to implement a given computational task. More specifically, how should these latent variables interact, and be embedded into the collective activity to produce the desired input-output transform? A promising approach for addressing this question is to exploit model reduction techniques to reverse-engineer large recurrent networks trained on the task using techniques inspired by artificial intelligence [88]. Such model-reduction approaches extract key latent variables and represent their interactions in an interpretable manner. One example of model reduction is model distillation, which consists of retraining networks of minimal size based on the activity of the larger RNN [89]. This technique has been recently used to study a hierarchical inference task [90] currently tested on mice [91]. Schaeffer et al [90] applied model distillation to synthesize an interpretable reduced model that solved the task using two latent variables which captured the low-dimensional dynamics of the fully trained RNN. A second example of model reduction is mean-field analysis, that can in particular be used to extract the dynamics of latent variables directly from trained RNNs [92]. Remarkably, these latent dynamics can be represented in terms of effective circuits that describe how incoming sensory latent variables interact with internally generated latent variables [86]. Using the mean-field approach, large RNNs trained on various tasks can be therefore reduced to effective circuits of latent variables, opening the door to a systematic mapping from computational tasks onto latent dynamics [92].

**Conclusion**

In this review, we put forward a specific definition of intrinsic and embedding dimensionality, and argue that they are potentially fruitful concepts for extracting computational principles from neural data. Testing this idea will require a systematic integration of non-linear analyses for dense neural recordings and computational models embodying computational and normative hypotheses, with the aim of determining the organization of latent variables underlying intrinsic dimensionality, and how they are put to use to implement computations.

**Acknowledgements**

MJ and SO were supported by the CRCNS project PIND funded through the National Institute of Health (NIMH: 1R01MH122025-01) and French Agence Nationale de la Recherche (ANR-19-NEUC-0001-01). SO was supported by the program "Ecoles Universitaires de Recherche" ANR-17-EURE-0017. MJ was supported by the Simons Foundation, the McKnight-Endowment Fund for Neuroscience and the McGovern Institute. SO thanks NeuroMatch Academy Pod 001 (Daniela Buchwald, Agustina Frechou, Habiba Noamany, Antonio Ortega) and Pod 162 (Elizaveta Kozlova, Viktoryia Kuryla, Anna Vasilevskaya, Egor Zverev) for discussions and reanalyses of V1 data (Fig. 1a). MJ thanks Nicholas Watters for discussions about linear decodability, factorization, and controllability in model networks.